\RequirePackage{fix-cm}
\documentclass[twocolumn]{svjour3}          
\smartqed  
\usepackage{graphicx}
\usepackage{amsfonts}
\usepackage[ruled,vlined]{algorithm2e}
\usepackage{caption}
\usepackage{amsmath}
\begin{document}

\title{Investigating the Effects of Robot Engagement Communication on Learning from Demonstration}

\author{Mingfei Sun, Zhenhui Peng, Meng Xia \and
        Xiaojuan Ma
}

\institute{Mingfei Sun, Zhenhui Peng, Meng Xia \and Xiaojuan Ma \at
              Department of CSE, Hong Kong University of Science and Technology, 
              \email{mingfei.sun@ust.hk}           
}

\date{Received: date / Accepted: date}

\maketitle

\begin{abstract}
Robot Learning from Demonstration (RLfD) is a technique for robots to derive policies from instructors' examples. 
Although the reciprocal effects of student engagement on teacher behavior are widely recognized in the educational community, it is unclear whether the same phenomenon holds true for RLfD. 
To fill this gap, we first design three types of robot engagement behavior (attention, imitation, and a hybrid of the two) based on the learning literature. 
We then conduct, in a simulation environment, a within-subject user study to investigate the impact of different robot engagement cues on humans compared to a "without-engagement" condition. 
Results suggest that engagement communication significantly changes the human's estimation of the robots' capability 
and significantly raises their expectation towards the learning outcomes, even though we do not run actual learning algorithms in the experiments. 
Moreover, imitation behavior affects humans more than attention does in all metrics, while their combination has the most profound influences on humans. 
We also find that communicating engagement via imitation or the combined behavior significantly improve humans' perception towards the quality of demonstrations, even if all demonstrations are of the same quality. 
\end{abstract}

\section{INTRODUCTION}
Robot Learning from Demonstration (RLfD) is a technique where a robot derives a mapping from states to actions, a.k.a \textit{policy}, 
from instructors' demonstrations~\cite{atkeson1997robot}. 
This technique has been shown to be successful in teaching robots physical skills by imitating instructors' body movements 
e.g., pole balancing\cite{atkeson1997robot}, tennis swings~\cite{ijspeert2002movement}, air hockey maneuvers~\cite{bentivegna2002humanoid}, etc. 
A standard RLfD process takes two steps: demonstration gathering step, which collects demonstrations from the human demonstrators, and policy deriving step, which reasons the underlying state-action mappings~\cite{argall2009survey}. 
Like a human learner, a robot in RLfD could have different strategies of gathering demonstrations according to its underlying policy derivation algorithms. For example, robots with the DAgger algorithm~\cite{ross2011reduction} learn progressively by taking incremental demonstrations from instructors, much like going through a scaffolding process~\cite{jackson1998design,saunders2006teaching}. A robot can also learn more proactively. For example, if equipped with Confidence-Based Autonomy (CBA)~\cite{chernova2009interactive}, an interactive algorithm for RLfD, a robot can request demonstrations at the states of which it has little or no knowledge. These learning strategies have been proven to be very effective and thus widely adopted in RLfD~\cite{laskey2017comparing}.

Unlike human learners, robots in previous RLfD processes rarely show any engagement cues during the learning process. They mostly remain stationary without giving any feedback, especially when instructors are giving demonstrations (i.e., in the demonstration gathering step). In human tutelage, engagement cues play an important role in shaping instructors' mental model of the learners~\cite{skinner1993motivation}. For example, learners' attentional engagement, e.g., gaze, indicates their points of interest in the instructions. Imitation, a behavioral engagement cue, shows learners' motivation to perform like the instructors~\cite{chernova2014robot}. It is reported that learner engagement cues could potentially affect instructor perceptions and behavior~\cite{guthrie2001classroom}. For example, in educational research, instructors are found to have the tendency to provide more support to learners of high behavioral engagement~\cite{skinner1993motivation}.

These effects of showing learning engagement, however, are less explored in the RLfD research, partly because designing engagement cues for robots in the context of RLfD is challenging. 
First, most of the existing methods for generating engagement cues in Human-Robot Interaction (HRI) cannot be directly applied to RLfD. 
For example, it is common practice in HRI to simulate robots' attentional engagement by directing their gaze towards visually salient elements (e.g., color or lightness~\cite{nagai2008toward}), specific objects (e.g., human faces~\cite{sidner2004look}) or predefined events (e.g., pointing gestures~\cite{breazeal2004humanoid}). 
This practice cannot be easily set up in RLfD because the robot's allocation of attention should follow the development of instructors' demonstrations. 
This is especially true in skill-oriented RLfD, where the robot needs to reproduce the body skills from the human demonstrator.
In this context, the attention should be subject to the demonstrations, i.e., body movements, which are less constrained and highly dynamic compared to a standard HRI process. 
Methods for generating other engagement cues, e.g., imitation~\cite{bailenson2005digital,riek2010my,li2015observer}, also need further adaptation to accommodate the dynamic nature of RLfD. 
Second, even if an engagement cue can be designed effectively, its deployment in RLfD should be in real-time with low computational cost.

To this end, we focus on skill-oriented RLfD and propose two novel methods (\textit{Instant attention} and \textit{Approximate imitation}) 
to enable robots to communicate their learning engagement in a RLfD process.
Note that we consider the demonstration gathering step as the interaction scenario since it determines the demonstration quality, which is crucial for the policy optimality~\cite{argall2009survey,sun2019adversarial}. 
We do not focus on designing effective learning algorithms for the demonstration learning. 
The learning engagement cues are generated as follows: the 
\textit{Instant attention} method generates robot attentional engagement by tracking instructors' body movements through particle filters; 
the \textit{Approximate imitation} method produces behavioral engagement, i.e., imitation, by partially mapping the instructor's joint movements to those of the robot with approximations. 
We then use the proposed methods to generate three modes of engagement communication (via attention, via imitation, and via a hybrid of the two) for robots in RLfD. 
To investigate the effects of the three engagement modes on humans,
we compare them with another mode (``without-engagement'' in which the robot remains stationary as most robots do in existing RLfD studies~\cite{atkeson1997robot,ijspeert2002movement,bentivegna2002humanoid}) by a within-subject user study in a simulation environment. 
Results suggest that robots with the proposed cues are perceived to be more engaged in the learning process and their behaviors are more socially acceptable in RLfD than the robots without. 
Also, having engagement cues significantly affect human's estimation of the robots' learning capabilities. 
The robots which communicate engagement in RLfD are perceived to be significantly more capable in learning than the robots without,
even though none of them are equipped with the learning algorithms.
The engagement communication also affects the human's expectation towards the final learning outcomes. 
Furthermore, behavioral cues influence humans' perceptions significantly more than attentional engagement does, 
while the hybrid cues significantly outperform the other two. 
We also find that showing behavioral or combined engagement significantly improves humans' evaluation of demonstration quality.
Specifically, the human participants perceived the demonstrations to be significantly more appropriate for the robot to learn when the robot communicates its engagement via behavioral or the mixed engagement, even though all demonstrations are actually of the same quality. 

The contributions of this paper are as follows. 
First, we propose two novel algorithms which allow robots to generate attention and imitation behavior to communicate its learning engagement with low computations in RLfD. 
Second, we developed a simulation platform to evaluate the effect of engagement communication in RLfD.
Third, we take a first step towards evaluating the effects of three types of engagement cues (attention, imitation, and hybrid) on humans. 
Through evaluation in a simulation environment with a humanoid robot learning the different skills from a simulated demonstrators,
we show interesting findings on the design of robot engagement communication in RLfD. 
To the best of our knowledge, this paper is the first to systematically investigate how the robot engagement communication affects the humans' perceptions and expectations of the robot in RLfD.

\section{Related work}\label{related-work}

\subsection{Robot Learning from Demonstration (RLfD)}

Robot Learning from Demonstration (RLfD) is also known as ``Programming by Demonstration'', ``imitation learning'', or ``teaching by showing''~\cite{schaal1997learning}.  Rather than exhaustively searching the entire policy space, RLfD enables robots to derive an optimal policy from demonstrators' (also called instructors) demonstrations~\cite{atkeson1997robot}. Usually, this technique does not require additional knowledge about programming and machine learning from human instructors, and thus opens up new possibilities for common users to teach robots~\cite{crick2011human}. Existing studies on RLfD focus mainly on policy derivation algorithms, e.g., mapping states to actions by supervised learning~\cite{chernova2009interactive}, updating the policy by value iteration in Reinforcement Learning~\cite{atkeson1997robot}, and recovering rewards to explain demonstrations by Inverse Reinforcement Learning~\cite{abbeel2004apprenticeship,sun2019adversarial}. 
Some studies also work on designing robots' reciprocal learning feedback to communicate what the robots have learned to human teachers, e.g., demonstrating the robot's current learned policy~\cite{calinon2007active}, providing verbal and/or nonverbal cues~\cite{koenig2010communication,admoni2015robot,pitsch2013robot,sun2017sensing,breazeal2004humanoid,chao2010transparent}, or visualizing where they succeed and fail~\cite{sena2018teaching}. 
These studies, however, largely overlook how the robots' engagement behavior would affect the instructors and their demonstrations, especially during the demonstration gathering step.
Hence, in this work, we consider how to generate behavior which allow robots to communicate their learning engagement to instructors, and investigate their potential effects on RLfD. 

\subsection{Engagement and learning engagement cues}
Engagement is a broad concept in HRI with many different definitions. Some studies focus on the whole spectrum of an interaction, and defines engagement as the process of initiating, maintaining, and terminating the interaction between humans and robots \cite{sidner2005explorations}. Others narrow the notion of engagement down to the maintenance of interactions, interpreting engagement as humans' willingness to stay in the interaction \cite{yamazaki2009revealing,szafir2012pay}. 

In the context of learning, engagement mainly refers to the state of being connected in the learning interaction, which can be measured from three aspects: cognition, behavior, and emotion~\cite{silpasuwanchai2016developing}. Cognitive engagement is closely related to the allocation of attention as it is one of the most important cognitive resources~\cite{pekrun2012academic}. Failure to attend to another person indicates a lack of interest~\cite{argyle1976gaze}. Thus, we adopt attention as a cue to communicate cognitive engagement in RLfD. Behavioral engagement is captured by task-related behavior, e.g., task attempts, efforts, active feedback, etc. Imitation, a common behavioral engagement signal, refers to ``non-conscious mimicry of the postures, mannerisms, facial expressions, (speech), and other behavior of one's interaction partners'' \cite{chartrand1999chameleon}. In interpersonal communications and HRI, the imitation behavior increases the likelihood of understanding~\cite{chartrand2005beyond}, interpersonal coordination~\cite{bernieri1991interpersonal} and emotional contagion~\cite{hatfield1993emotional}. In the context of learning, the imitation behavior also indicates the robot's internal status in learning, e.g., the progress and motivation~\cite{chernova2014robot}. Thus, we use imitation as a way to communicate the behavioral engagement for robots in RLfD. Emotional engagement is associated with the affective status evoked by the interaction, including valence and arousal. Despite its importance, emotional engagement is hard to apply in RLfD since most existing RLfD robot systems lack the full ability to express emotions. In the scope of this paper, we define the robot learning engagement as the involvement in the learning process, with a focus on its cognitive engagement, i.e., attention, and behavioral engagement, i.e., imitation. The following subsection presents related work on generating attention and imitation behavior to communicate engagement. 

\subsection{Robots' communication of engagement}
In HRI, a robot can communicate its \textbf{attention} via different physical channels, e.g., gaze~\cite{lockerd2004tutelage,kuno2007museum,mutlu2006storytelling,mutlu2009footing}, head orientation~\cite{lockerd2004tutelage,sun2017sensing}, and body postures~\cite{takayama2011expressing}. Regardless of which channel they use, robots are usually programmed to pay attention to salient elements, including but not limited to colors \cite{breazeal2004humanoid}, objects with visual intensity \cite{nagai2008toward}, and movements \cite{breazeal2004humanoid,nagai2008toward}. For example, Nagai~\emph{et al.} regarded visually outstanding points in the surroundings, in terms of their colors, intensities, orientations, brightness and movements, as points of attention~\cite{nagai2008toward}. Other work directs robots' attention to specific objects, e.g., human faces~\cite{sidner2004look} and colorful balls~\cite{anzalone2015evaluating} to name a few, or predefined events, e.g., pointing gestures~\cite{lockerd2004tutelage}. For example, Sidner~\emph{et al.} designed a robot that pays attention to participants' faces for most of the time~\cite{sidner2004look}. Lockerd~\emph{et al.} drove the robot attention mechanism with interaction events, such as looking at an object when it is pointed at or looking at a subject when the person takes a conversational turn\cite{lockerd2004tutelage}. To accommodate multiple events, a state transition diagram is usually adopted to control any attention shifts~\cite{breazeal2004humanoid,lockerd2004tutelage}. Though these studies provide insightful information about the design of robot attention,  their approaches may not easily be applicable to skill-oriented RLfD as the point for attention in instructors' body movements is dynamically changing.

Compared to attention, the \textbf{imitation} behavior has been less widely adopted as a robot engagement cue. The robot imitation of a human participant's behavior in real-time is inherently challenging due to the correspondence problems~\cite{argall2009survey} as well as the robot's physical constraints~\cite{kim2009stable,suleiman2008human,koenemann2014real}. Hence, instead of generating full-body imitation behavior, some HRI researchers proposed to do partial imitations. For example, Baileson and Yee built an immersive virtual agent that subtly mimicked people's head movements in real-time~\cite{bailenson2005digital}. A similar imitation strategy was applied by Riek~\emph{et al.} to a chimpanzee robot~\cite{riek2010my}. In addition to head imitation, gesture ``mirroring'' has also been implemented by Li~\emph{et al.} on a robot confederate~\cite{li2015observer}. Although these studies showed that partial imitation behavior improve participants' perception of robots' capabilities~\cite{gonsior2011improving,fuente2015influence}, they mainly used ruled-based methods~\cite{bailenson2005digital} or predefined behavior~\cite{li2015observer}, which may not be transferable to RLfD scenarios. In this work, we employ the same strategy and allow robot learners to make partial imitations. Different from existing work, we take an algorithmic approach to automatically generating approximate imitations of instructors' body movements for robots in real-time.

\section{LEARNING ENGAGEMENT MODELING}
This sections presents two methods for generating engagement cues. The first subsection briefly introduces human body poses and forms the basis of the proposed methods. The remaining subsections describe the methods in detail. 


\subsection{Representation of the body pose}\label{body-pose-descrip}
\begin{figure}
\centering
\includegraphics[width=0.8\linewidth]{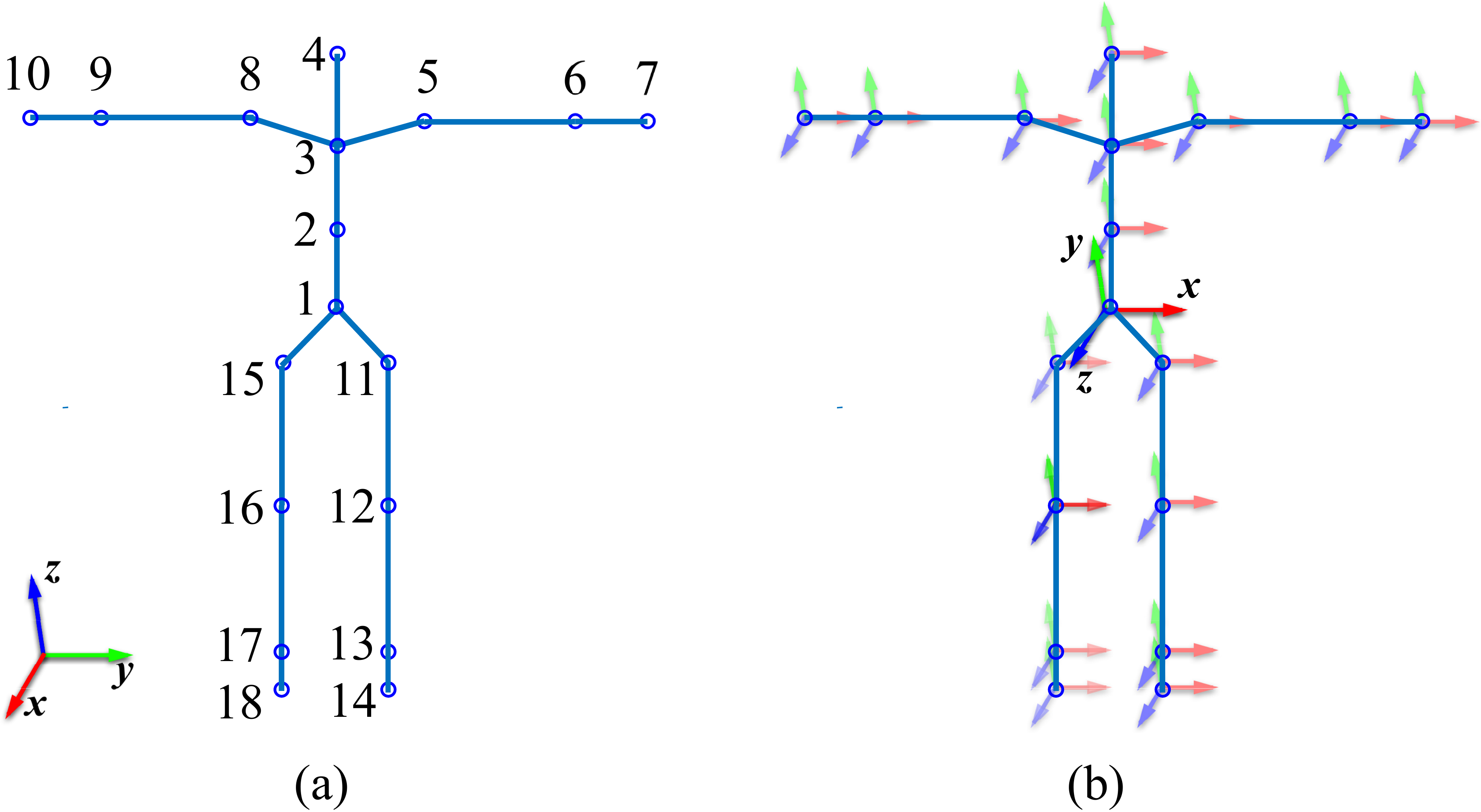}
\caption{(a) A body pose in the \textbf{position form}: all joints are described in a single frame by their positions; (b) A body pose in the \textbf{transformation form}: each joint has its own frame and the skeleton defines parent-child structures and translations between frames; the frame with $x$-$y$-$z$ labels is the root frame and is referred in the sensor frame.}
\label{fig:pose}
\end{figure}

In RLfD, instructors usually demonstrate a point via their body movements. Our proposed methods thus use human body poses to generate attention and imitation behavior. A body pose is usually depicted by a tree-like skeleton, with nodes as human joints and links as body bones (shown in Figure~\ref{fig:pose}). Mathematically, this skeletal structure can be represented in two forms~\footnote{Usually, the two forms are readily available in most body pose extracting sensors, e.g., Kinect.}: \textit{the position form} and \textit{the transformation form}. 

\textbf{Position form:} The position form describes the body pose in a single frame of reference (usually the sensor frame), as shown in Figure~\ref{fig:pose}(a). In this form, the pose skeleton is denoted as $[J^{(1)}, J^{(2)}, ..., J^{(n)}]$, where $J^{(i)} \in \mathbb{R}^{3}$ is the position vector of the $i$-th joint in the skeleton, and $n$ is the number of joints. This form gives for each joint its global position, providing the potential attention point for the robot. Hence it is used for the \textit{Instant attention} algorithm to generate robot attention points. 

\textbf{Transformation form:} The transformation form describes the body pose in a series of frames of reference~\cite{sun2019estimating}, as shown in Figure~\ref{fig:pose}(b). In particular, each joint has its own frame (a right-handed frame), and the links in the tree-like skeleton define parent-child structures between frames. 
The pose of a non-root joint is then described by a translation (i.e., the bone length) and a rotation (i.e., joint movement) in its parental frame, with the root joint (often the hip joint) described in the sensor frame. This form decomposes a human body movement into joint rotations (body-independent) and joint translations (body-dependent) in a way that the movement can be easily imitated by robots: just mapping the rotations onto robot joints. We denote this form as $[T_1, T_2, T_3, ..., T_n]$, and use it for the \textit{Approximate imitation} algorithm to obtain approximate imitation behavior.

\subsection{Instant attention}



The attentional engagement for robots is generated based on the cognitive theories on human attention. Generally speaking, a generation process of human visual attention involves two stages~\cite{jonides1983further}: first, attention is distributed uniformly over the visual scene of interest; then, it is concentrated to a specific area (i.e., it is focused) for gaining information~\cite{eriksen1972temporal}. In a skill-oriented RLfD process, the instructor demonstrates skills mainly through their body joint poses. The above mechanism thus corresponds to that the human joints of interest are tracked uniformly at the initial stage, and then one joint providing the most information for learning is picked as an attention point. As for a demonstration learning, the more predictable/track-able a body joint movement is, the less information the robot could gain from that part, and consequently, less attention the robot should pay to it. In other words, if a body joint moves out of expectation the most among all joints, it will be worth paying attention to.

To this end, we use the particle filter (PF) as it is robust and effective in predictions~\cite{konidaris2012robot} and tracking~\cite{arulampalam2002tutorial}. In short, PF is a Bayesian filter which uses a group of samples to approximate the true distribution of a state~\cite{thrun2005probabilistic}.
Particularly, given the state observations, PF employs many samples (called \textit{particles}) to describe the possible distribution of that state. The particles are denoted as
\begin{equation}
X_{t} := x_{t}^{[1]}, x_{t}^{[2]}, ..., x_{t}^{[M]}
\end{equation}
Here $M$ is the number of particles in the particle set $X_{t}$. Each particle $x_{t}^{[m]}$ (with $1 \leq m \leq M$) is a hypothesis as to what the true state might be at time $t$, and is first produced by a prediction model $p(x_t | z_{1:t-1})$ which is based on all history observations $z_{1:t-1}$, i.e., $x_{t}^{[m]} \sim p(x_{t} | z_{1:t})$. 
At each updating stage, particle $x_{t}^{[m]}$ is then re-sampled according to the \textit{importance weight} $w_{t}^{[m]}$, i.e., the probability that the the particle $x_{t}^{[m]}$ is consistent with the current observation $z_{t}$, i.e., $w_{t}^{[m]} = p(z_{t} | x_{t}^{[m]})$.
In other words, each $x_t^{[m]}$ survives into the next stage with the probability $w_{t}^{[m]}$. For more details on the particle filter, refer to~\cite{thrun2005probabilistic}.

We apply one PF to track each relevant joint during the human demonstration. Specifically, state $\mathbf{x}_{t}^{[m]} \in \mathbb{R}^3$ describes the joint position in the sensor frame. We assume the state transits with additive Gaussian noise:
\begin{equation}\label{equ:state-trans}
\mathbf{x}_{t}^{[m]} \sim \mathbf{x}_{t-1}^{[m]} + \Delta_{t-1} + \mathcal{N}\big(\textbf{0}, \sigma_{t} \mathbf{I} \big)
\end{equation}
where $\Delta_{t-1}$ is the observed joint shift: $\Delta_{t-1} = J_{t-1} - J_{t-2}$; and $\mathcal{N}\big( \textbf{0}, \sigma_{t} \mathbf{I} \big)$ is the multivariate normal distribution with zero mean and diagonal covariance matrix $\sigma_{t} \mathbf{I}$. The \textit{importance factor} for each particle is defined to be exponential to the Euclidean distance between the predicted and observed joint position:
\begin{equation}\label{equ:likelihood-adjust}
w_{t}^{[m]} = \eta e^{-2 \big( \mathbf{x}_{t}^{[m]} - J_{t} \big)^{T} \big( \mathbf{x}_{t}^{[m]} - J_{t} \big)}
\end{equation}
where $\eta$ is the normalizer. Each joint in the body pose is tracked by a \textit{particle cloud}, a group of particles $\mathbf{X}_{t}$. In order to dynamically adjust the cloud size in accordance with the joint movement, the variance $\sigma_{t}$ is set to be proportional to the average Euclidean distance between the predicted and observed joint position:
\begin{equation}\label{equ:distribution-adjust}
\sigma_{t} =  \frac{\alpha}{M} \sum_{m} \big[ \big( \mathbf{x}_{t}^{[m]} - J_{t} \big)^{T} \big( \mathbf{x}_{t}^{[m]} - J_{t} \big) \big]
\end{equation}
where $\alpha$ is a hyper-parameter and $M$ is the number of particles. The $\sigma_{t}$ indicates the cloud size: the greater the $\sigma_{t}$ is, the more attention the robot should pay to the associated joint. 
Thus, the joint with maximum $\sigma_{t}$ corresponds to the attention point. 
In the experiment, the $\alpha$ is set to $0.02$ for best tracking of human joints. 
Note that, though the importance factor is calculated as the distance between predicted and observed joint positions, it is not equivalent to the measure of joint acceleration. 
In particular, the predicted joint position is just an estimate, and the difference between predicted and observed joint positions measures how much the estimate deviates from the truth. 
The importance factor thus implies the unpredictability and can only be computed after the current observation is available.

\begin{figure}
\centering
\includegraphics[width=0.8\linewidth]{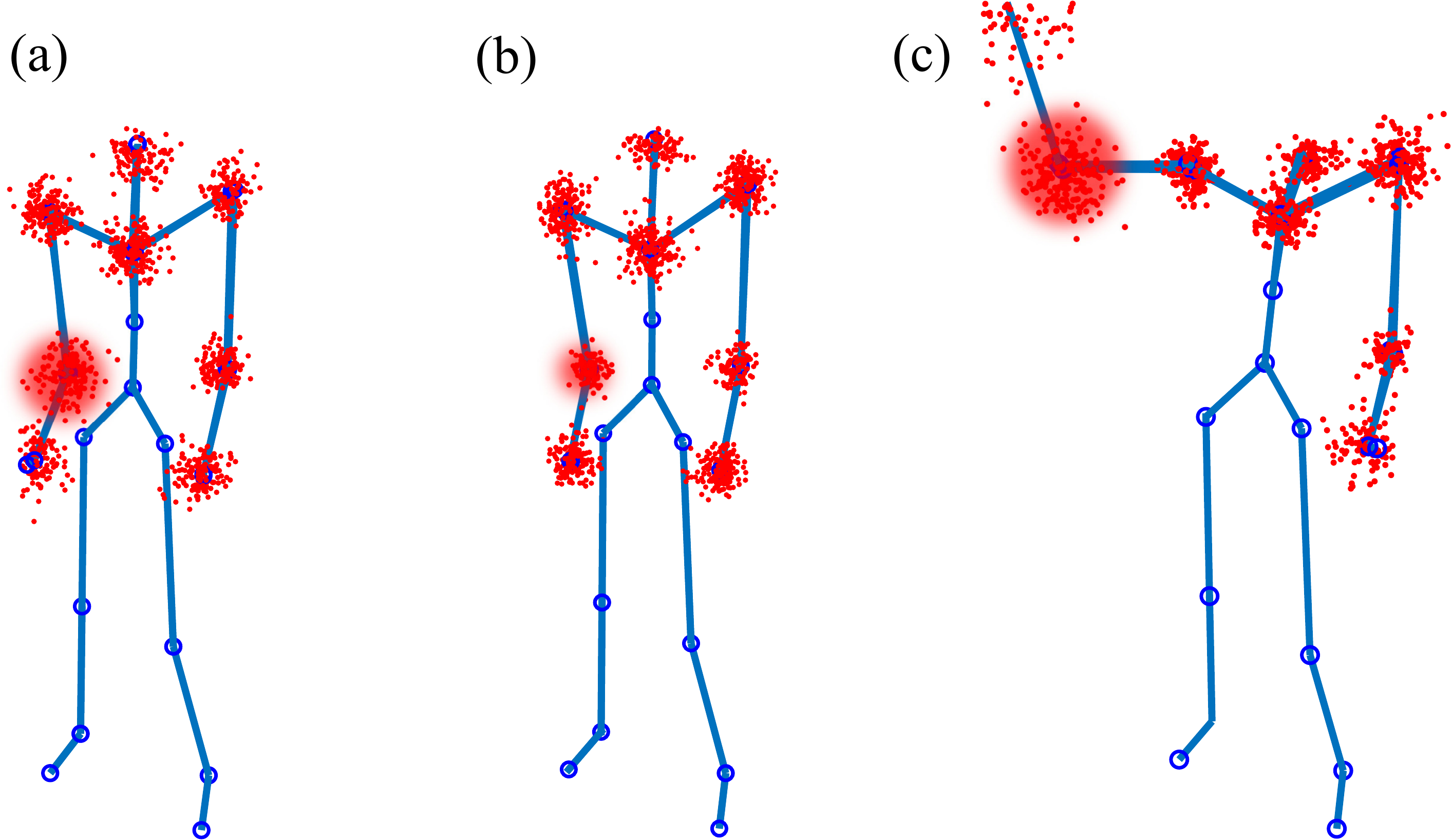}
\caption{The particle clouds evolve over time: (a) all clouds are initialized at the same size; (b) if the joint movement is small, the cloud shrinks: the picked cloud becomes smaller since the elbow did not move; (c) if the joint moves out of its cloud region, the cloud grows to catch the movement: the picked cloud becomes larger to adapt to the elbow's movements.}
\label{fig:joint-cloud}
\end{figure}

\begin{figure}
\centering
\includegraphics[width=1.0\linewidth]{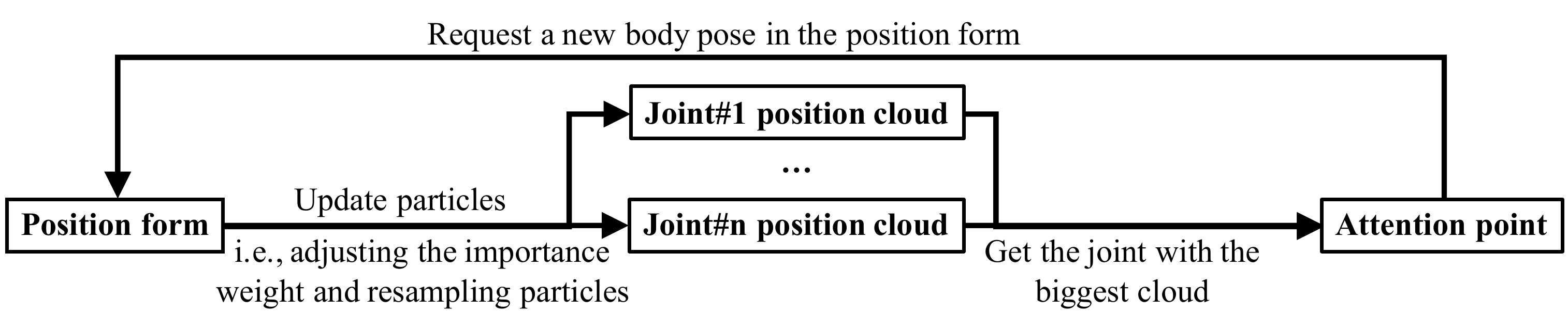}
\caption{The flow chart of the \textit{Instant attention}.}
\label{fig:flowchart-instant-attention}
\end{figure}

Figure~\ref{fig:joint-cloud} illustrates how the PF works to generate attention. The particle cloud functions as the robot's prediction of the joint future movements, and is subject to change based on the current observations. Initially, the robot predicts the movements of all body joints of interest to be the same, i.e., all clouds are of the same sizes. During a demonstration, when a joint moves out of its cloud region, beyond the robot's prediction, the cloud grows to catch that movement and the robot will thereafter be likely to pay attention to that joint. Likewise, if the joint movement is small, within the robot's prediction, or no movement at all, the cloud shrinks and chances are small that the attention will be given to that joint. Overall, the cloud size indicates the predictability of the instructor's body movements as well as the level of attention the robot needs to pay. At each time, the joint with the biggest cloud is picked as the attention point. This process loops with every new body pose as shown in Figure~\ref{fig:flowchart-instant-attention}. 

We now present a practical algorithm for \textit{Instant attention} to generate attentional engagement instantly for robots (Algorithm~\ref{alg:attention}). The algorithm takes \textit{TrackedJoints} $JSet_{\text{tracked}}$ and the \textit{BodyPose} in the position form $[J_{t}^{(1)}, J_{t}^{(2)}, ..., J_{t}^{(n)}]$ as input, and outputs one attention point at each time. Specifically, the \textit{TrackedJoints} contains the joints required to be tracked. In practice, the joints to be tracked are task-dependent, and should be defined according to the possible attention points on the instructor's body. 
For example, a cooking robot may only need to track the instructor's upper body movements and the joint correspondence can be configured by the developers based on the robot physical structures. 
Another input \textit{BodyPose} is the human body pose in the position form. The algorithm runs as follows: first, it initializes a particle filter with the same covariance for each tracked joint (line 2-4). Then it estimates the distribution of the next joint position (line 9-11), followed by the estimation correction given the current position observations (line 12-13). Finally, the algorithm adjusts the covariance of the noise distribution to capture the joint movement (line 14), and the attention point is found by selecting the joint with the maximum covariance value (line 15). 

\begin{algorithm}[t]
\SetNlSty{texttt}{[}{]}
\SetAlgoNlRelativeSize{0}
\LinesNumbered
\KwIn{TrackedJointSet ${JSet}_{\text{tracked}}$;
BodyPose $\big[ J_{t}^{(1)}, J_{t}^{(2)}, ..., J_{t}^{(n)} \big]$, where $J_{t}^{(i)}$ is the 3D position of $i$-th joint at time $t$}
\KwOut{AttentionPoint $P_{\text{a}} \in \mathbb{R}^3$}
\Begin{
  \For{ each joint $j$ in ${JSet}_{\text{tracked}}$}{ 
    initialize $j$-th particle filter for joint $j$\;
    initialize $\sigma_{t}^{(j)} = 1$ for joint $j$\;
  }
  \For{each joint $J_{t}^{(i)}$ in $\big[ J_{t}^{(1)}, J_{t}^{(2)}, ..., J_{t}^{(n)} \big]$ }{
    \If{$J_{t}^{(i)}$ is in ${JSet}_{\text{tracked}}$}{
      $\Delta_{t-1}^{(i)} = J_{t-1}^{(i)} - J_{t-2}^{(i)}$\;
      obtain particles $\mathbf{x}_{t-1}$ from $i$-th particle filter\;
      \For{m = 1 to M }{
        sample $\mathbf{x}_{t}^{[m]}$ with probability $\propto \mathbf{x}_{t-1}^{[m]} + \Delta_{t-1}^{(i)} + \mathcal{N}\big(\textbf{0}, \sigma_{t} \mathbf{I} \big)$ \tcc*[r]{Equation\ref{equ:state-trans}}
        calculate $w_{t}^{[m]} = \eta e^{-2 \big( \mathbf{x}_{t}^{[m]} - J_{t} \big)^{T} \big( \mathbf{x}_{t}^{[m]} - J_{t} \big)}$ \tcc*[r]{Equation\ref{equ:likelihood-adjust}}
      }
      \For{m = 1 to M }{
        draw new particles $\mathbf{x}_{t}^{[m]}$ with probability $\propto$ $w_{t}^{[m]}$\;
      }
      update $\sigma_{t}^{(i)} = \alpha/M * \sum_{m} \big[ \big( \mathbf{x}_{t}^{[m]} - J_{t} \big)^{T} \big( \mathbf{x}_{t}^{[m]} - J_{t} \big) \big]$ \tcc*[r]{Equation\ref{equ:distribution-adjust}}
    }
  }
  $P_{a} = \text{argmax}_{J_{t}^{(i)}}\sigma_{t}^{(i)} $  \;
  \Return $P_{a}$ \;
}
\caption{Instant attention}
\label{alg:attention}
\end{algorithm}

\begin{figure}
\centering
\includegraphics[width=0.8\linewidth]{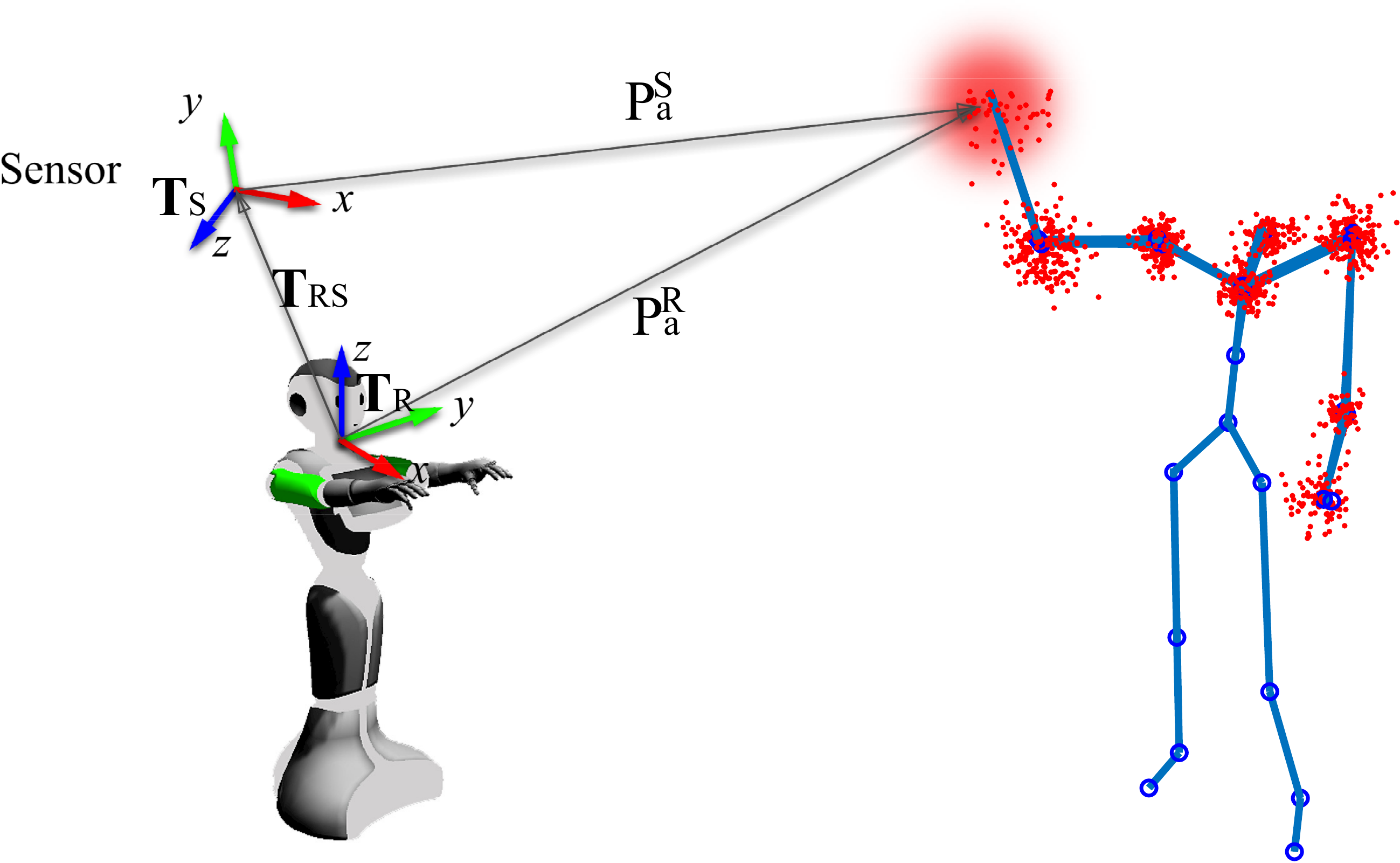}
\caption{The attention point $P_{a}^{S}$ is located in the sensor frame $T_{S}$. We need to do the transformation $P_{a}^{R} = T_{RS}P_{a}^{S}$ to get $P_{a}^{R}$ in the robot head frame $T_{R}$, where $T_{RS}$ is the transformation from $T_{S}$ to $T_{R}$.}
\label{fig:attention-tf}
\end{figure}

Once an attention point is generated, say $P_a$, it is worth mentioning that $P_a$ is actually located in the sensor frame. In order to obtain the accurate attention point of the robot, a further transformation is required. Figure~\ref{fig:attention-tf} illustrates how to transform $P_a$ in the sensor frame $T_S$ into the robot head frame $T_R$ given the transformation $T_{RS}$ from $T_S$ to $T_R$. 

The \textit{Instant attention} method has several advantages. First, unlike other mechanisms (salience-based, object-based or event-based), this method utilizes the particle cloud to track the instructor's joint movements, and automatically produces attention points based on the information gained from the movements. Second, the attention point is generated and shifted smoothly because the spatial size of the cloud evolves smoothly. Specifically, the particle distribution $p(X_{t})$ is iteratively sampled based on their previous distribution $p(X_{t-1})$ by the \textit{importance weight} $w_t$, i.e., a $p(x_{t-1}^{[m]})$ in $X_{t-1}$ survives into $X_{t}$ with probability $w_t^{[m]}$, even if the joint moves abruptly (i.e., $\mathbf{x}_{t}^{[i]} - J_t$ is large). Also, the cloud is immune to noises and outliers, e.g., joint vibrations caused by sensors, since small turbulence (no matter the exact speed) will not change cloud size (the $\sigma_t$ is averaged over all predicted states), while existing speed-/spatial-position-based methods could cause gaze jerks or sudden gaze shifts due to these noise/outliers. Third, the joints to be tracked can be dynamically changed, offering a flexible and adjustable attention mechanism based on the RLfD task. 

\subsection{Approximate imitation}

Behavior imitation in robotics is usually formulated as an optimization problem, which needs to find the joint correspondence first~\cite{argall2009survey}, and then solve the inverse kinematics for the robot structure~\cite{grochow2004style}. Both of the processes are difficult, computationally intensive, and robot-configuration-dependent, hence not applicable for generating imitation behavior for general robots. On the other hand, psychological results reported that people mimic behavior to communicate engagement by adopting similar postures or showing similar body configurations according to the context~\cite{chartrand1999chameleon}. We thus relax the behavior imitation in robotics as follows: First, the robot is not required to search blindly for the best joint correspondence since the joint correspondence is task-dependent. We allow the user to explicitly specify the joint correspondence according to the RLfD context. 
Second, for those robot joints whose Degree of Freedom (DoF) do not match the human joint, we only set the joint angles for the available robot joints to approximate the human movements. Though this solution of approximation may not be optimal in the sense of behavior mimicry, it runs very fast (in real-time) to generate behavioral engagement, achieving a balance between simplicity and optimality.

\begin{figure}
\centering
\includegraphics[width=1.0\linewidth]{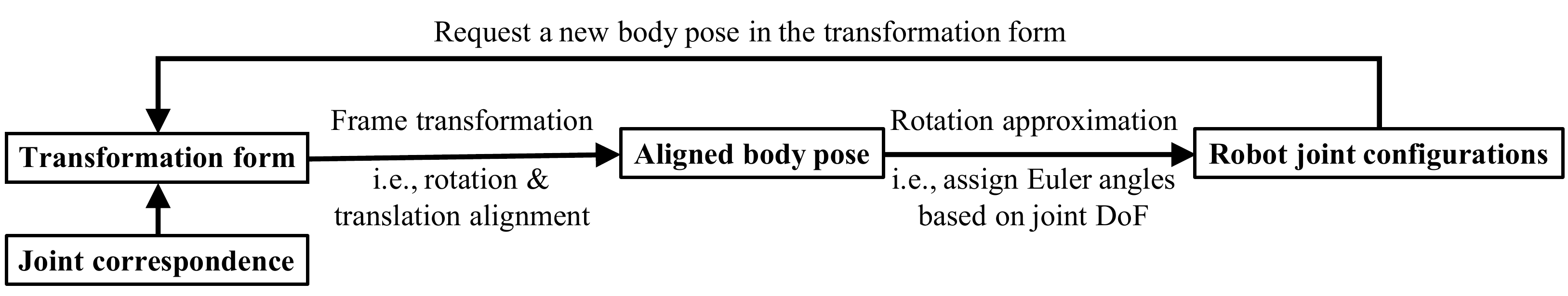}
\caption{The flow chart of the \textit{Approximate imitation}}
\label{fig:flowchart-approx-imitate}
\end{figure}

To achieve this, we propose the algorithm \textit{Approximate imitation}, which allows robots to generate similar motions as the demonstrator's for specified joints. Given the joint correspondence, the algorithm runs with two steps: frame transformation, and rotation approximation, as presented in Figure~\ref{fig:flowchart-approx-imitate}. 

\begin{figure}[]
    \centering
    \includegraphics[width=1.0\linewidth]{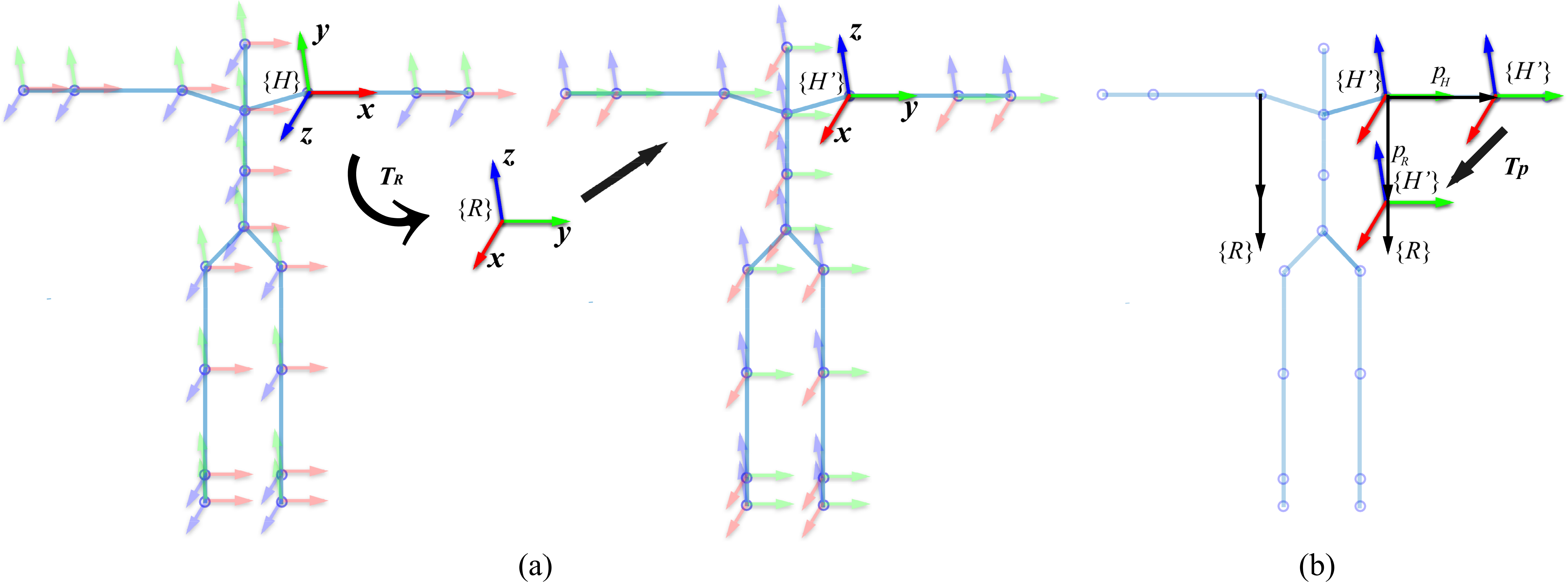}
    \caption{Frame transformation. (a) Rotation alignment: aligning the local frame $\{H\}$ of the human body pose with the corresponding robot joint frame $\{R\}$ by rotation matrix $R$. The aligned local frame is $\{H'\}$. (b) Translation alignment: translating $\{H'\}$ in its parent frame by $T_p$ to match the corresponding robot frame $\{R\}$ so that the human pose link $p_H$ is aligned with the robot link $p_R$.}
    \label{fig:frame_alignment}
\end{figure}

The frame transformation is to transform the instructor's body pose to match the robot frames. To be specific, we leverage the transformation form of body poses to decompose the frame matching into two steps: first, rotation alignment, and then translation alignment. The rotation alignment is to rotate the human joint frames so that their axes are aligned with the robot joint frames, as shown in Figure~\ref{fig:frame_alignment}(a); the translation alignment is to translate the human joint frames in their parent frames so that the initial skeletal structure of the demonstrator's body matches the robot initial configurations, as shown in Figure~\ref{fig:frame_alignment}(b). To sum up, we represent the rotation alignment as $T_R$ in the joint frame, $\{H\}$, and the translation alignment as $T_p$ in the parent frame of $\{H\}$, $\{H_p\}$ (both represented in Homogeneous transformation). Then for $\{H\}$, its frame transformation is $T_{H}^{H_p}T_p\{H\}T_R$, where $T_{H}^{H_p}$ is the transformation from $\{H_p\}$ to $\{H\}$.

\begin{figure}
\centering
\includegraphics[width=0.9\linewidth]{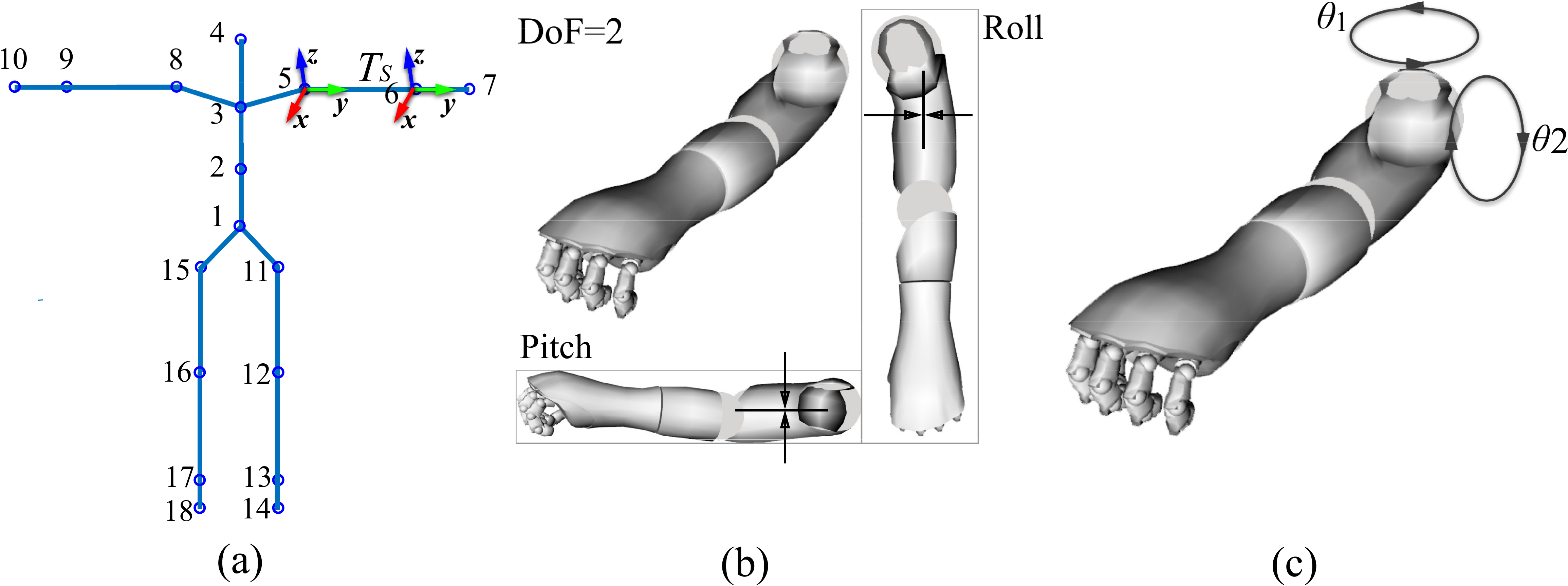}
\caption[The LOF caption]{Rotation approximation: (a) the instructor's left shoulder has a DoF of 3 and its transformation is $T_S$ ; (b) the robot shoulder joint has a DoF of 2: \textit{roll} and \textit{pitch}; (c) the robot rotates for its shoulder the roll joint with $\theta_{roll}$ and then the pitch joint with $\theta_{pitch}$, without considering $\theta_{yaw}$.}
\label{fig:tf-config}
\end{figure}

Since the DoF of the robot joint may not equal the DoF of its corresponding human joint, we could not have the exact movement mapping. Instead, we use the robot joint to approximate the human joint rotations as follows. First, a human joint rotation is converted into Euler forms, $(\theta_{\text{roll}}, \theta_{\text{pitch}}, \theta_{\text{yaw}})$. Second, if the DoF of a robot joint is 3 (\textit{roll}, \textit{pitch} and \textit{yaw}) and exactly matches the human DoF, then the conversion is straightforward: rotate for the robot joint with \textit{roll} first, then \textit{pitch}, and finally \textit{yaw}. If the DoF of a robot joint is 2 (e.g., \textit{roll} and \textit{pitch}), then the conversion can be approximated as rotating with \textit{roll} first, and then \textit{pitch}. If the DoF of a robot joint is 1 (e.g., \textit{roll} only), then rotate with \textit{roll} only. For example, in Figure~\ref{fig:tf-config}, the robot arm has the same structure as the demonstrator's but with different joint DoF, as shown in Figure~\ref{fig:tf-config}(a) and (b). It can approximate the instructor's left arm movement by first converting $T_S$ (the rotation) into Euler angles $(\theta_{roll}, \theta_{pitch}, \theta_{yaw})$, and then setting the joint roll to $\theta_{roll}$, and the joint pitch to $\theta_{pitch}$ for the shoulder, ignoring the $\theta_{yaw}$, as shown in Figure~\ref{fig:tf-config}(c).

We now present the algorithm \textit{Approximate imitation} in Algorithm~\ref{alg:approximate-imitate}. The algorithm takes joint correspondence \textit{JointCorrespondence}, and instructor's body pose \textit{JointMovement} in transformation form as input, and outputs the joint configurations, \textit{JointConfigs}, for the robot. Specifically, \textit{JointCorrespondence} defines the joint mapping, $\{J^H_i \rightarrow J^R_i\}$, from human joint $J^H_i$ to robot joint $J^R_i$ for part joints. The \textit{JointMovement} is represented as a series of transformations along the skeletal structure, $[T_1, T_2, ..., T_n]$ (see Section~\ref{body-pose-descrip} for more details). The algorithm runs as follows: first, it calculates the frame transformations from $J^H$ to $J^R$, and saves the rotation alignment and translation alignment in $Rotation\_align$ and $Translation\_align$ (line 3-5). Then for each joint movement $T_i$ in $[T_1, T_2, ..., T_n]$, the algorithm transforms it into the corresponding robot frame $T_i^\prime$ by translation and rotation alignment, followed by a conversion into the Euler form (line 7-8). The algorithm proceeds by selecting the right angles from $\theta_{roll}$, $\theta_{pitch}$, and $\theta_{yaw}$ for the robot joint according to the DoF of the robot joint (line 9-16). The joint configurations are saved in $q_R$, and returned as the final output. 

\begin{algorithm}[t]
\SetNlSty{texttt}{[}{]}
\SetAlgoNlRelativeSize{0}
\LinesNumbered
\KwIn{JointCorrespondence $\{J^H_1 \rightarrow J^R_1, J^H_2 \rightarrow J^R_2, ..., J^H_n \rightarrow J^R_n\}$, JointMovement=$[T_1, T_2, ..., T_n]$}
\KwOut{JointConfigs $q_R$}
\Begin{
  $q_R$ = []; $Rotate\_align = []$; $Translate\_align = []$ \;
  \For{ $(J^H_i \rightarrow J^R_i)$ in $\{J^H_1 \rightarrow J^R_1, J^H_2 \rightarrow J^R_2, ..., J^H_n \rightarrow J^R_n\}$}{
    $Rotate\_align$.append(rotateAlign($J^R_i$, $J^H_i$)) \;
    $Translate\_align$.append(translateAlign($J^R_i$, $J^H_i$)) \;
  }
  \For{$i$ in $[1, 2, ..., n]$}{
    $T_i^\prime$ = $Translate\_align[i] * T_i * Rotate\_align[i]$ \;
    $(\theta_{roll}, \theta_{pitch}, \theta_{yaw})$ = convertToEuler($T_i^\prime$) \;
    \eIf{$DoF(J^R_i) == 3$}{
        append $[\theta_{roll}, \theta_{pitch}, \theta_{yaw}]$ to $q_R$ \;
    }{
        \eIf{$DoF(J^R_i) == 2$}{
            append $[\theta_{roll}, \theta_{pitch}]$ to $q_R$ \;
        }{
            \If{$DoF(J^R_i) == 1$}{
                append $[\theta_{roll}]$ to $q_R$ \;
            }
        }
    }
  }
  \Return $q_R$ \;
}
\caption{Approximate imitation}
\label{alg:approximate-imitate}
\end{algorithm}

The \textit{Approximate imitation} method has several advantages for generating imitation behavior for robots in RLfD. First, this algorithm runs in real-time as the imitation is only partially taken place on the instructor's body poses. In particular, we take advantage of local transformations of body poses to avoid solving inverse kinematics for the whole robot joints, which is computationally intensive and may also not have the closed form solutions. Also, instead of finding the exact mapping for robot joint angles, we set configurations based on the DoF of the robot joint to achieve a similar motion trend. This conversion may sometimes distort movements, but, still, the directions and trends are captured (as reflected in \ref{sec:evaluation}). Second, this method is generic and applicable to standard skill-oriented RLfD. Depending on the RLfD scenario, we can also assign different joint correspondences to do a partial imitation.
For other types of RLfD, e.g., object-related demonstrations or goal-oriented learning from demonstrations, we can also apply the proposed method to generate the approximate imitation based on the object or the goal.
Specifically, we can replace the joint transformations with the poses of the object or the goal, and generate the target $\theta_{roll}$, $\theta_{pitch}$, and $\theta_{yaw}$. 
Then we can adopt the inverse kinematic solvers to calculate a set of joint configurations to move the robot's end-site to the target pose $(\theta_{roll}, \theta_{pitch}, \theta_{yaw})$. 
Based on the DoF and the space constraints of the robot end-effectors, we can make the similar approximations to have the end-effector only achieve the $roll$ pose, the $roll$ and $pitch$ pose, or the complete target pose.

\section{Evaluation}\label{sec:evaluation}
This section first introduces our RLfD simulation platform, then describes a preliminary study for determining the timing of imitating behavior, and finally presents the main user study. 

\subsection{RLfD simulation platform}
\begin{figure}
\centering
\includegraphics[width=1.0\linewidth]{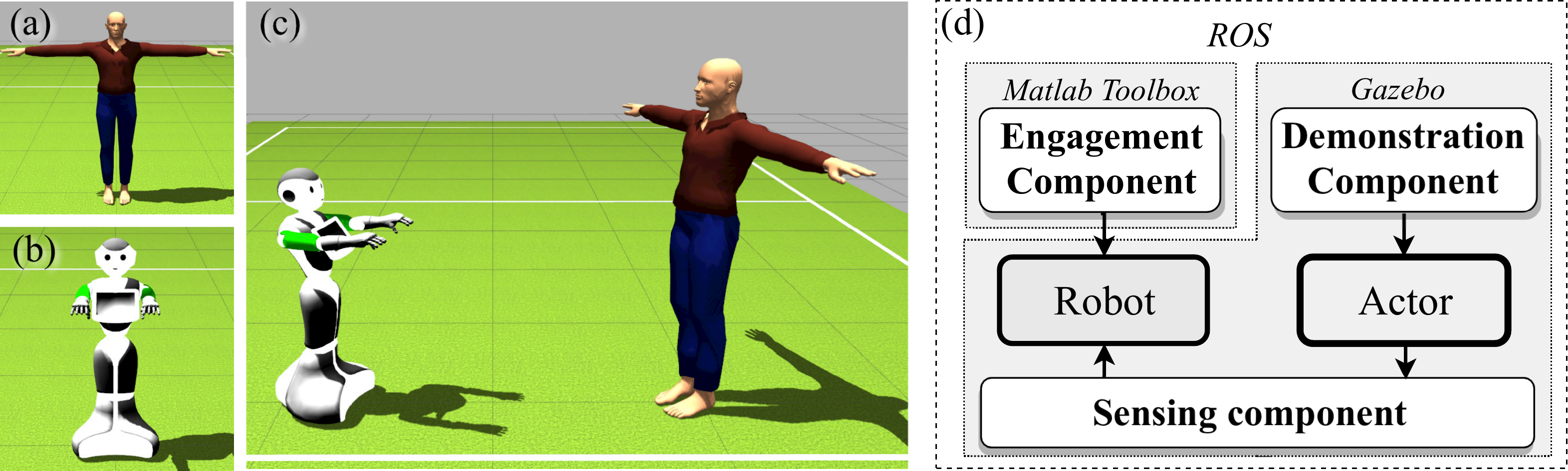}
\caption{RLfD simulation platform: (a) the simulated human instructor; (b) the virtual Pepper robot; (c) the instructor and robot are facing towards each other for teaching and learning; (d) platform composition.}
\label{fig:sim-env}
\end{figure}

Our RLfD simulation platform is composed of a virtual human instructor and a robot, as shown in Figure~\ref{fig:sim-env}(a) and (b). The virtual human instructor performs different yet controlled types of movement skills, while the robot (a Pepper) needs to capture motion and learn skills from the instructor. Both parties stand facing each other in a simulated 3D space, as shown in Figure~\ref{fig:sim-env}(c). 

\begin{figure}
\centering
\includegraphics[width=1.0\linewidth]{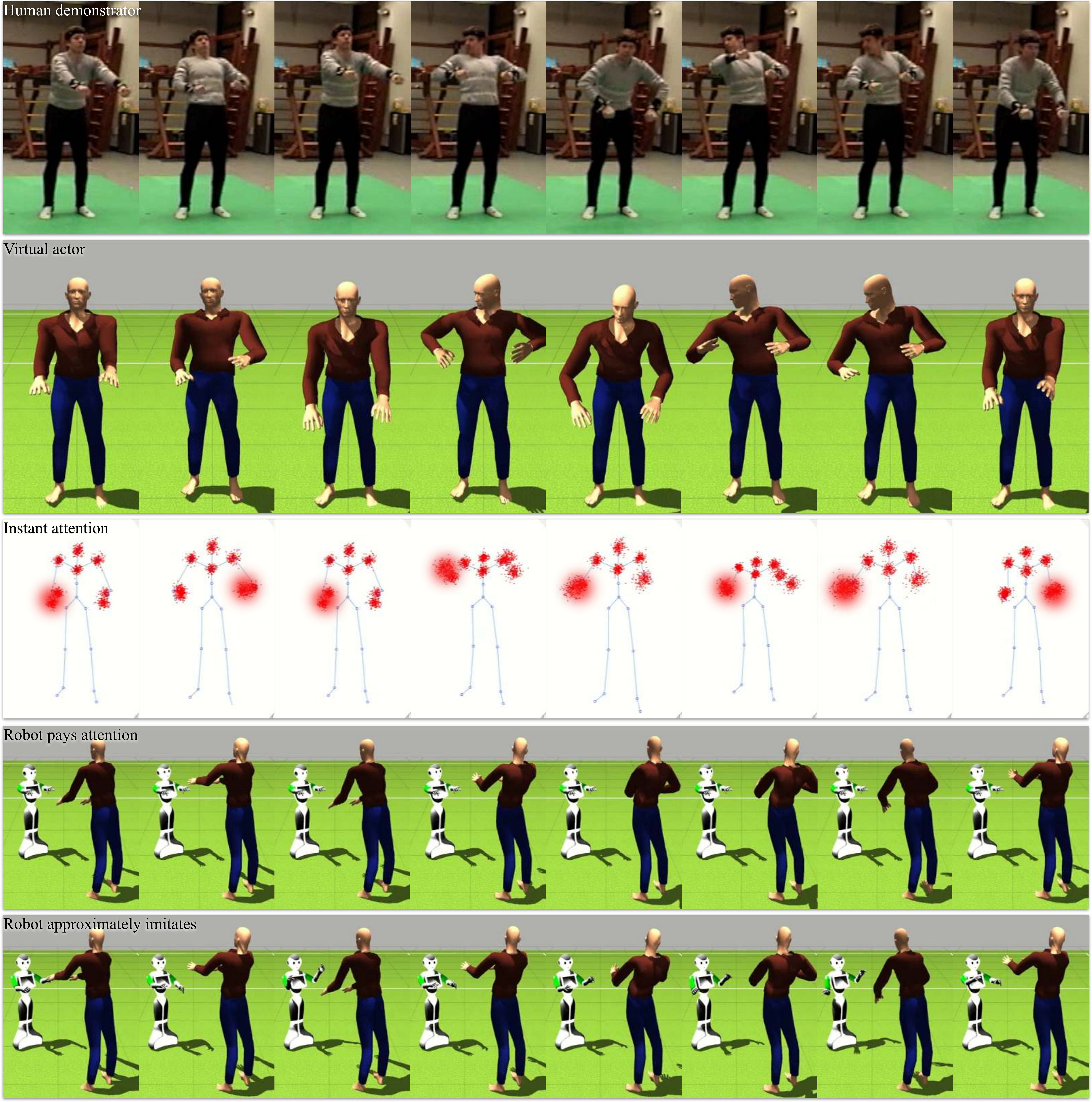}
\caption{An example to show how the platform works}
\label{fig:sys-test}
\end{figure}

The simulation platform has three major components: demonstration component, sensing component, and engagement component, as shown in Figure~\ref{fig:sim-env}(d). The demonstration component determines what movements the instructor needs to perform. We exploit motion capture (MoCap) data to simulate real movements. The MoCap data are recorded by 3D motion capturing systems with high precision, and are usually used for simulations and animations~\cite{gleicher1998retargetting}. The sensing component serves as a pose sensor, extracting body poses from the virtual instructor. This component also converts body poses between two representations (global positions and local transformations). Finally, the engagement component controls the robot's engagement communication. Based on the proposed algorithms, the robot could choose one of the three ways to communicating engagement in RLfD: showing attention (\textit{A-mode}), showing imitation (\textit{I-mode}), and showing both (\textit{AI-mode}). We further add one more mode, i.e., no engagement (\textit{N-mode}), to evaluate the effectiveness of these three modes. 
In \textit{N-mode}, the robot just stands near the instructor and remains stationary without any body movements. 
Compared with the \textit{A-mode}, the robot's gaze is fixed on the demonstrator's face and is not affected by the demonstrator's body movements.

In this simulated RLfD, the tasks for robots to learn are sports skills performed by a virtual instructor. We chose sports skills for robots to learn as this type of movement has often been adopted in RLfD~\cite{ijspeert2002movement,bentivegna2002humanoid}. Four types of sports movements, i.e., boxing, rowing, swimming, and frisbeeing, are selected from CMU Graphics Lab Motion Capture Database\footnote{http://mocap.cs.cmu.edu/} as these four sports involve movements of various body parts. 
Regarding the policy deriving algorithms, even the state-of-the-art method may fail to deliver good learning outcomes, which may in turn change their perception towards the demonstration gathering. 
Thus, to minimize any side-effects or biases introduced by the performance of the learning algorithms, we do not utilize any learning algorithms, and the robot has no actual learning ability in the demonstration gathering process. 
In the other words, the robot only communicates its engagement when observing the human demonstrations by showing different cues and will not learn the sport skills in the following experiments and studies. 

Figure~\ref{fig:sys-test} presents an example of how the simulation platform works. The first row shows the human instructor's real demonstration, which is then re-targeted onto the instructor, as shown in the second row. The third and forth rows present the running of \textit{Instant attention} and robot showing attention (\textit{A-mode}). The last row presents the approximate imitation behavior of the robot (\textit{I-mode}). We purposely rotate the 3D scene in the last two rows to get a better view of robot communicating engagement.

We chose online simulation rather than a field test due to the following constraints and concerns: First, due to the current limitations of RLfD techniques, the demonstrators are usually required to wear motion-capture devices, confined in a designated space, and repeatedly showcase the target movements. This could potentially impact on their interaction with robots and perception of the robot behavior. Also, limited by physical abilities, robots, e.g., Pepper, barely move without making undesirable noises, jerks, and vibrations, which could disturb the human participants and influence their assessment of robot learning. We thus use simulation in our experiment to avoid all these side effects and unexpected outcomes. Furthermore, we purposely select a viewpoint that allows the participants to have a better view of both the robot's and the instructor's behavior, i.e., the staging effect~\cite{thomas1995illusion}. Second, the robot's engagement behavior could be evaluated in a more consistent and repeatable manner in a simulation. In a field test, the instructor's demonstrations are usually non-repeatable and could be easily influenced by robots' reactions. The simulation allows different engagement cues to be compared without bias. Second, the simulation provides a controllable and measurable environment to monitor and evaluate a system's performance from various perspectives, which is often a necessity before algorithms are deployed in RLfD.

This simulation platform was built upon the Gazebo simulator \footnote{http://gazebosim.org/} and the Robot Operating System (ROS). We use the Matlab Robotics System Toolbox \footnote{https://www.mathworks.com/hardware-support/robot-operating-system.html} to facilitate the algorithm implementation.

\subsection{Preliminary study}
In interpersonal communication, a person's imitation behavior, also called \textit{mirroring behavior}, often happens after the partner's target behavior with certain time delay~\cite{chartrand1999chameleon,hove2009s}. 
In this paper, we generate such mirroring behavior via the approximation mechanism. 
We need to determine the exact time delay so that users can correctly recognize imitation as a learning engagement cue.
We run a within-subject pilot experiment to check the appropriate timing of robot imitation relative to the target action.

\textbf{Manipulated variable.} We set time delay as the independent variable in this study and experiment with three intervals: $0.5s$, $1.0s$, and $2.0s$. Technically, we used a buffer to store instructors' body poses to postpone any imitation behavior. After proper setup, the buffer size was set to $15$, $30$, and $60$ to achieve an appropriate time delay of about $0.5s$, $1.0s$, and $2.0s$, respectively.

\textbf{Subject allocation.} We recruited 30 participants (mean age: 35.5, female: 12) via Amazon Mechanical Turk (AMT) who had no prior experience with physical or virtual robots. Each participant watched three simulated RLfD videos corresponding to the three delay intervals. In the videos, the instructor was teaching the robot some type of sports skill, and we staged the 3D scene at a fixed angle for a better view of the robot imitations. We counterbalanced the presentation order of the different time delays.

\textbf{Dependent variables.} Participants watched videos showing the robot imitating the instructor with three different time delays. They were informed that the robot is supposed to learn sports skills from the demonstrator. After each video, they were asked to rate their agreement on a 7-point Likert scale as to whether the robot in the video is actually learning. 

\begin{figure}
\centering
\includegraphics[width=1.0\linewidth]{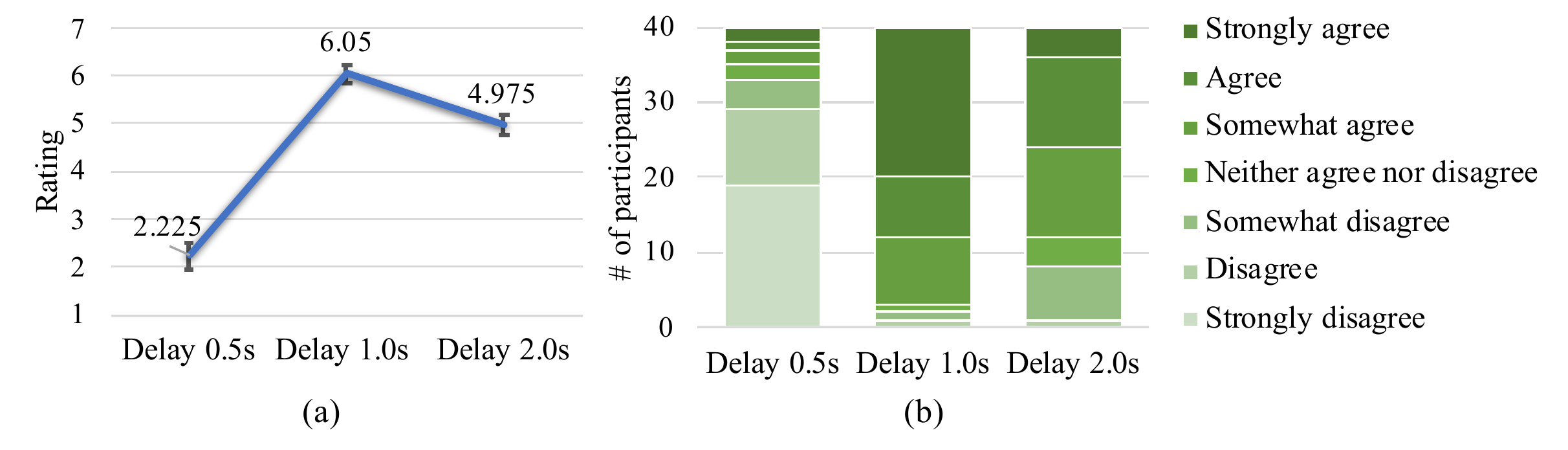}
\caption{Results for timing: (a) average ratings; (b) rating distribution.}
\label{fig:delay}
\end{figure}

Figure~\ref{fig:delay} presents the average and overall rating distribution on different time delays.  
We run a repeated measures ANOVA with time delay as the factor, and find that there is a significant difference in delay-induced perception of robot learning engagement ($F(2, 58)=88.37, p<0.01, \eta^2 = .76$).
Results of the Bonferroni post-hoc test suggest that the engagement rating of delaying for $1.0s$ is significantly higher than that of delaying for $0.5s$ ($p<0.01$) and $2.0s$ ($p<0.01$). 
Overall, setting the imitation time delay to $1.0s$ can effectively communicate robots' learning engagement ( 70\% agree and strongly agree). 
We apply this configuration to the \textit{Approximate imitation} algorithm in the main user study. 

One might be wondering that why the rating difference between $0.5s$ and $1s$ delay is noticeably dramatic,
even larger than the difference between $1s$ and $2s$ delay. 
The cause may possibly be the approximation mechanism adopted for generating the mirroring behavior. 
When the delay time is small (e.g., $0.5s$), the approximate imitation algorithm generates the movement in a very responsive manner, 
almost at the same pace with the demonstrator's movement. 
The subjects are likely to feel that the robot is showing, rather than following, the demonstrator's movement. 
As the delay time becomes longer (e.g., $1s$), the movement following effects becomes more obvious, and the robot appears to be learning from the demonstrator by mimicking his/her behavior. 
Consequently, the ratings between the $0.5s$ and $1s$ in terms of robot communicating learning engagement become higher. 
Such dramatic rating difference also confirms the necessity and importance of using the preliminary study to determine the appropriate delay time for the followed studies.

\subsection{Main study}
To evaluate the effectiveness of engagement communication and our proposed cues on participants' perception of the robot and the demonstration, we conducted a within-subject experiment on an RLfD simulation platform, with an additional "without engagement" condition (\textit{N-Mode}) as the baseline. 

\subsubsection{Hypothesis} 

Our proposed methods generates different types of engagement cues for robots to express their engagement. Accordingly, we first hypothesize that:

\textbf{\textit{H1}}. 
\textbf{1)} Regardless of actual cues taken, robots that communicate engagement are perceived to be significantly more engaged (\textit{H1a}) in learning, and their learning behavior is significantly more socially acceptable (\textit{H1b}) than those in the \textit{N-mode}. 
Further, \textbf{2)} 
imitation cue will receive a significantly higher engagement rating than attention cue (\textit{H1c}), while combined cues will be rated significantly the most (\textit{H1d}). 
Similarly, \textbf{3)} 
imitation cue will be rated significantly more acceptable than attention cue (\textit{H1e}) while combined cues will be rated significantly the most (\textit{H1f}).

According to educational theory postulating that learners' engagement cues, especially behavioral engagement, could have reciprocal effects on instructors~\cite{skinner1993motivation}, we hypothesize that:

\textbf{\textit{H2}}. Robots communicating engagement via different cues will have significantly different influences on human participants.  
Specifically, \textbf{1)} regardless of the cues, communicating engagement will significantly influence humans' estimation of the robot learning capability (\textit{H2a}), and significantly raise the humans' expectations towards the learning outcomes (\textit{H2b}) than no communication.
Further, \textbf{2)} 
imitation cues will lead to a significantly higher estimation of the robot's capabilities than attention cues (\textit{H2c}) while combined cues have the most significant influence than others. (\textit{H2d}). 
Similarly, \textbf{3)}
imitation cues will result in a significantly higher expectation towards the learning outcome than attention cues (\textit{H2e}) while combined cues have significantly the highest expectation than others (\textit{H2f}).

We further hypothesize that the robot showing different engagement behavior can affect humans' assessment of demonstration quality. More specifically:

\textbf{\textit{H3}}. \textbf{1)} Regardless of the exact demonstrations shown to robots, different engagement cues will influence the human participants' assessment of the demonstration quality. Specifically, demonstrations for robots with attention cues, imitation cues and the hybrid cues will be rated as significantly more appropriate (in terms of the expected robot capabilities) than that without engagement cues even if they are actually the same (\textit{H3a}). Further, \textbf{2)} demonstrations for robots with imitation cues and the hybrid cues will have a significantly higher rating in appropriateness than that with attention cues (\textit{H3b}). 

In the study, these different aspects were measured via post-study questions with 7-point Likert scale answers, as shown in Figure~\ref{fig:engagement} and Figure~\ref{fig:perception}. 
We derived these questions in the user study based on the previous research on Human-Robot Interactions and robot learning. 
Specifically, the questions to measure robot communicating engagement are adapted from the engagement studies~\cite{Strait:2015:TMH:2702123.2702415,Sun:2017:SHE:3025453.3025469}; 
the questions to measure participants' expectation towards the robot learning capability are derived based on the studies on human expectations and assessment of human-robot collaborations~\cite{Kwon:2018:ERI:3171221.3171276}. 
We also took two steps to ensure the effectiveness of the answers to all the questions, . 
First, the questions could only be answered after participants took necessary actions to understand the experiment. 
For example, the questions to measure engagement were only visible when the participants finished watching the full learning videos; 
and the questions to measure the participants' expectation also require the participants to provide the answers and their reasons (those without giving reasons could not proceed to next questions). 
Second, all answers were manually checked to reject any invalid responses, e.g., a response with the same answers to all questions, and a response with vague and inconsistent comments. 

\subsubsection{User study design}

The study consisted of five sessions: one introductory session and four experimental sessions. The introductory session requested demographic information and presented a background story to engage users: the participant has a robot team of four for an Olympic game. They needed to assess the robots' performance when they were under a professional coach's tutelage. In experiment sessions, participants watched the human instructor's movements first and then monitored the robot learning process in the RLfD simulation platform. After each session, participants were required to fill post-study questionnaires. 
Each session checked one mode, and modes were counter-balanced with learning tasks. 
Specifically, we randomized the order of engagement modes and the four physical skills to ensure the mode applies evenly across different skills and the skill also occurs evenly across different modes. 
We recruited 48 participants (mean age: 30.9, female: 6, no prior experiences with teaching robots and no participation in the preliminary study) from Amazon Mechanical Turk (AMT). 

During the experiment, we asked the participants to rate if they perceived the robot was paying attention or imitating based on its behavior. 
This served as the manipulation check for validity, ensuring that our designs indeed convey the intended type of engagement.

\subsubsection{Analysis and results}

\textbf{Manipulation check}.
The manipulation check for different engagement communications shows that the manipulation is effective (for attention cue: repeated measures ANOVA, $F(3, 141) = 153.79, p < 0.01, \eta^2 = .80$; for imitation cue: repeated measures ANOVA, $F(3, 141) = 197.45, p < 0.01, \eta^2 = .84$). Robots in \textit{A-mode} ($M=5.53, SD=1.85$) and \textit{AI-mode} ($M=6.17, SD=1.11$) are indeed perceived to show more attention than robots in \textit{N-mode} ($M=2.53, SD=1.83$); Bonferroni post-hoc test $p<0.05$. Also, more imitation behavior is reported by subjects with robots in \textit{I-mode} ($M=4.98, SD=1.33$) and \textit{AI-mode} ($M=6.05, SD=1.22$) than robots in \textit{N-mode} ($M=1.88, SD=1.57$); Bonferroni post-hoc test $p<0.05$.

\begin{figure}
\centering
\includegraphics[width=1.0\linewidth]{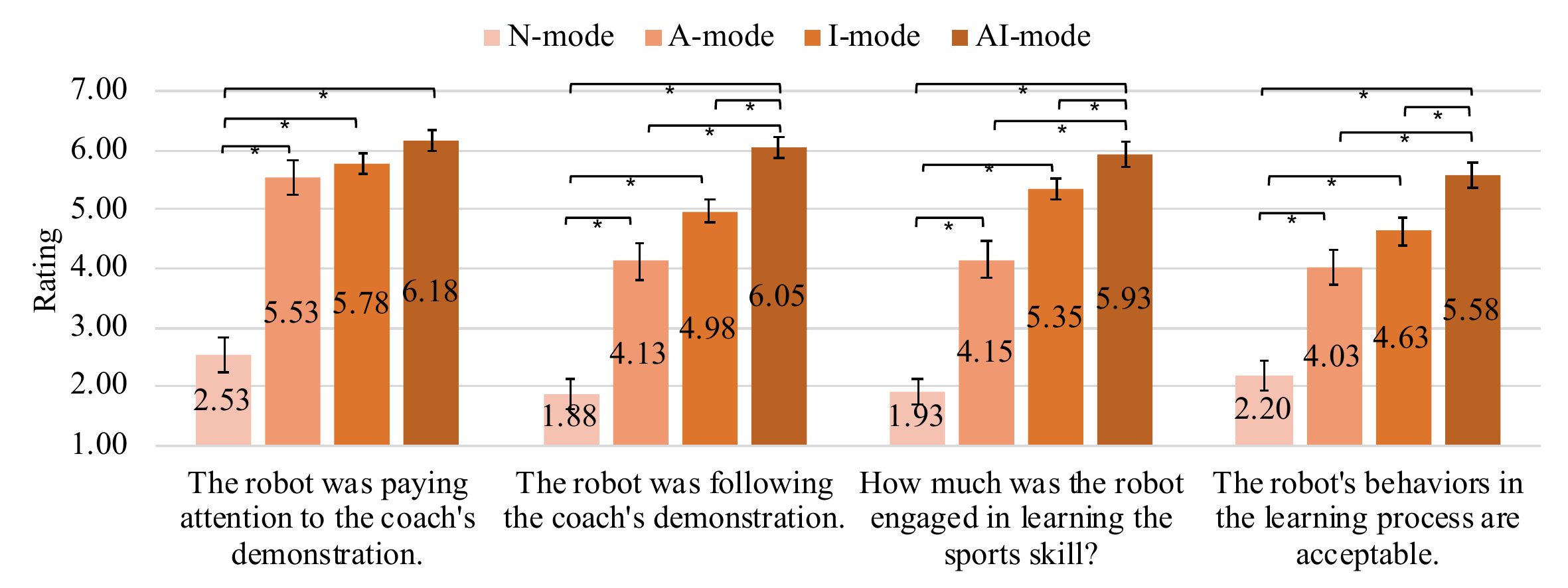}
\caption{Ratings on robot engagement communications and their behavior in RLfD.}
\label{fig:engagement}
\end{figure}

\textbf{Efficacy of proposed engagement cues}. 

We analyze participants' ratings via a one-way repeated measures ANOVA with the mode as the independent variable. We find that both attention and imitation cues significantly improve the ratings of robots' engagement levels and their behavior, as shown in Figure~\ref{fig:engagement}. Specifically, the robots with \textit{A-mode} ($M=5.53, SD=1.85$), \textit{I-mode} ($M=5.78, SD=1.03$) and \textit{AI-mode} ($M=6.17, SD=1.11$) are perceived to be significantly more engaged in the learning process than the robot in \textit{N-mode} ($M=2.53, SD=1.83$); repeated measures ANOVA, $F(3, 141) = 153.79, p < 0.01, \eta^2 = .80$, \textbf{\emph{H1a}} accepted. Consequently, subjects accept the robots' behavior in RLfD (\textit{A-mode}: $M=4.02, SD=1.78$, \textit{I-mode}: $M=4.62, SD=1.44$, and \textit{AI-mode}: $M=5.58, SD=1.39$) significantly more than the robot in \textit{N-mode} ($M=2.20, SD=1.60$); repeated measures ANOVA, $F(3, 141) = 102.89, p < 0.01, \eta^2 = .73$, \textbf{\emph{H1b}} accepted. Further, in terms of engagement, combined cues are reported to be significantly better than single cues; Bonferroni post-hoc test $p < 0.01$; \textbf{\emph{H1d}} accepted. in terms of acceptability, combined cues are reported to be significantly better than single cues; Bonferroni post-hoc test $p < 0.01$; \textbf{\emph{H1f}} accepted. However, we do not notice a significant difference between imitation cue and attention cue, thus \textbf{\emph{H1c}} and \textbf{\emph{H1e}} are both rejected. Therefore, \textbf{\emph{H1}} is partially accepted.

Based on these analyses, we therefore conclude that:

\begin{quotation}
\noindent \emph{Overall, our results partially support \textbf{\textit{H1}}: showing attention, imitation or both are perceived to be significantly more engaged in learning, and is significantly more acceptable. Also, showing both behavior is perceived to be significantly better than showing only one behavior. However, no significant difference can be found between showing attention and showing imitation. }
\end{quotation}

\begin{figure}
\centering
\includegraphics[width=1.0\linewidth]{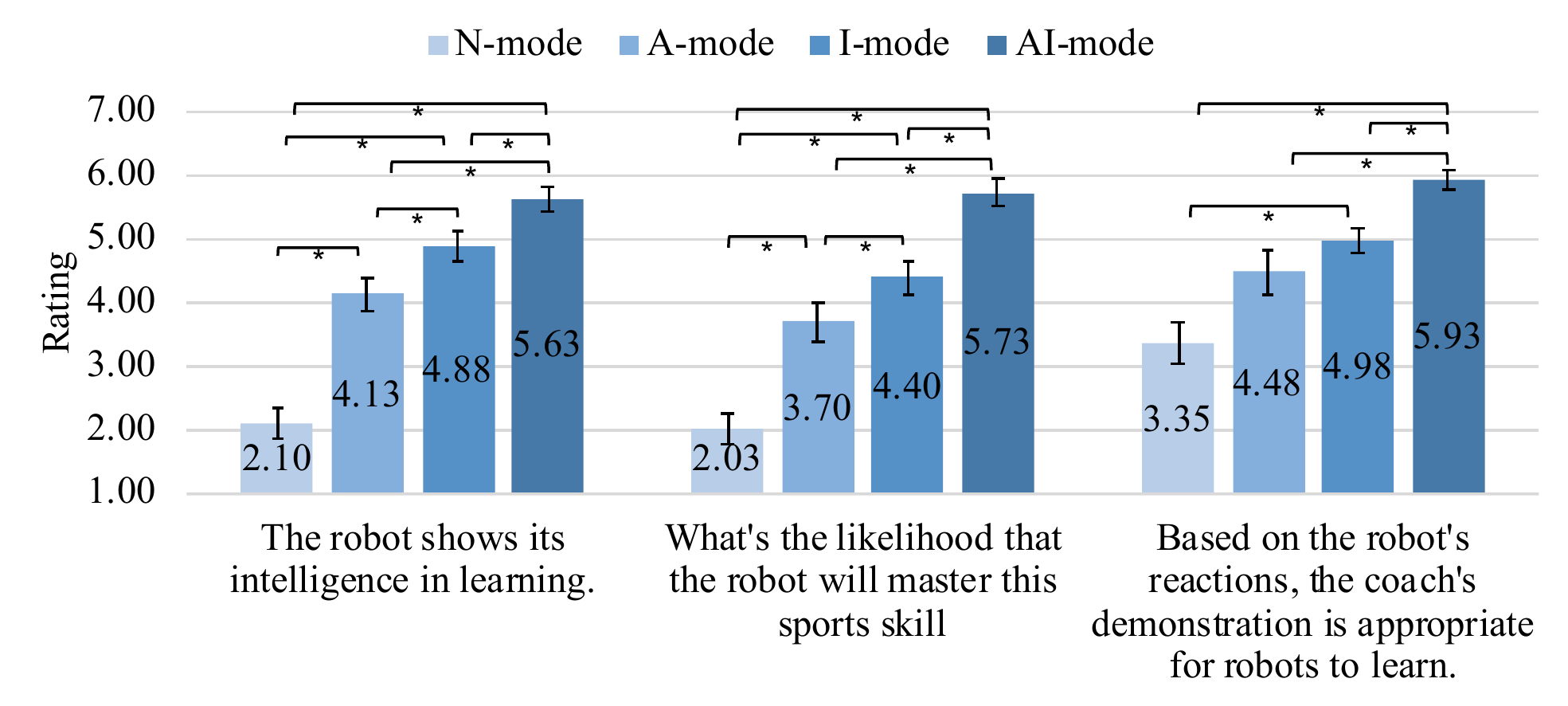}
\caption{Ratings on the effects of engagement communication on the participants' perception and their assessment of demonstration qualities.}
\label{fig:perception}
\end{figure}

\textbf{Effects of engagement cues on participants' perception}. 

We then compare the effects of different engagement cues on subjects' perception via a one-way repeated measures ANOVA with the mode as the independent variable. 
In general, robot engagement communication significantly enhances the participants' estimation of robots' learning capabilities and the participants' expectation of the learning outcomes, even if none of the robots in the experiment have the learning ability (no learning algorithms are adopted in the user study). 
Specifically, in terms of estimating the robots learning capability, participants rated the robots in \textit{A-mode} ($M=4.13, SD=1.70$), \textit{I-mode} ($M=4.88, SD=1.49$) and \textit{AI-mode} ($M=5.63, SD=1.21$) to be significantly more intelligent than the robots in \textit{N-mode} ($M=2.10, SD=1.45$); repeated measures ANOVA, $F(3, 141) = 155.25, p < 0.01, \eta^2 = .80$; \textbf{\emph{H2a}} accepted. 
Similarly, participants rated the robots with engagement behavior (\textit{A-mode}: $M=3.70, SD=1.94$, \textit{I-mode}: $M=4.40, SD=1.63$, and \textit{AI-mode}: $M=5.73, SD=1.47$) to be more likely to master the skills than the robots without (\textit{N-mode}: $M=2.02, SD=1.59$); repeated measures ANOVA, $F(3, 141) = 125.38, p < 0.01, \\
\eta^2 = .76$; \textbf{\textit{H2b}} accepted. 

In addition, showing behavioral engagement, i.e., \textit{I-mode}, have significantly more influences on the participants than showing attentional engagement, i.e., \textit{A-mode}. 
In particular, the robots in \textit{I-mode} ($M= 4.88, \\ SD=1.49$) are perceived to be significantly more capable of learning the demonstrated skills than the robots in \textit{A-mode} ($M= 4.13,  SD=1.70$); repeated measures ANOVA, $F(3, 141) = 155.25, p < 0.01, \eta^2 = .80$; \textbf{\emph{H2c}} accepted.
Similarly, the robots in \textit{I-mode} ($M= 4.40, SD=1.63$) receive significantly higher ratings than the robots in \textit{A-mode} ($M= 3.70, SD=1.94$) in terms of participants' expectation towards the learning outcomes; repeated measures ANOVA, $F(3, 141) = 125.38, p < 0.01, \eta^2 = .76$. Thus, \textbf{\textit{H2e}} accepted.  

Further, we also notice significant differences between robots in \textit{AI-mode} and robots in other modes. 
Specifically, robots in \textit{AI-mode} show significantly more intelligence in learning ($M=5.63, SD=1.21$) than robots in \textit{N-mode} ($M=2.10, SD=1.45$), \textit{A-mode} ($M=4.13, SD=1.70$), and \textit{I-mode} ($M=4.88, SD=1.49$); repeated measures ANOVA, $F(3, 141) = 155.25, \\ p < 0.01, \eta^2 = .80$; \textbf{\emph{H2d}} accepted. 
Also, the robots in \textit{AI-mode} ($M=5.73, SD=1.47$) are estimated by the participants to be significantly more likely to master the skill than the robots in modes ( \textit{N-mode}: $M=2.02, SD=1.59$, \textit{A-mode}: $M=3.70, SD=1.94$ and \textit{I-mode}: $M=4.40, SD=1.63$); \textbf{\textit{H2f}} accepted. Note that in all different engagement modes and different skill settings, the robots are equipped with no learning algorithms and thus have no actual learning abilities. 

\begin{quotation}
\noindent \emph{Overall, our results support \textbf{\textit{H2}}: communicating engagement significantly influence the humans' estimation of the robots' learning capabilities, and significantly changes their expectation towards the final learning outcomes, even though none of the robots have the learning abilities. 
Moreover, the behavioral engagement in RLfD, i.e., imitation, presents significantly more influence on the participants than the attentional engagement. 
Furthermore, communicating engagement via two cues at the same time have significantly more effects on participants than communicating engagement via a single cue.}
\end{quotation}

\textbf{Effects on participants' assessment of demonstration qualities}. 

Finally, we analyze the participants' ratings on the appropriateness of instructors' demonstrations. 
As shown in Figure~\ref{fig:perception}, no significant difference can be found between \textit{A-mode} ($M= 4.48, SD=2.10$) and \textit{N-mode} ($M= 3.35, SD=2.08$); \textbf{\textit{H3a}} rejected. 
However, compared with \textit{A-mode}, only \textit{AI-mode} ($M= 5.93, SD=1.00$) significantly improves the participants' assessment of demonstration quality in RLfD, Bonferroni post-hoc test $p < 0.01$; \textbf{\textit{H3b}} partially accepted. 
Note that in different engagement modes, the skills to be learned are all generated by the same set of MoCap data. Thus all demonstrations are actually of the same quality. 

\begin{quotation}
\noindent \emph{Overall, our results partially support \textbf{\textit{H3}}: communicating behavioral engagement or combined engagement will significantly improve participants' assessment of demonstration qualities, while showing attention cannot, even though all the demonstrations are actually of the same quality.}
\end{quotation}
 
Further, in the comments collected from the user study, we found that most participants explicitly stated that the robots without behavioral engagement may fail in learning, 
and accordingly, they were more likely to adjust future demonstrations when the robots communicated no engagement or only attentional engagement. 

\section{Discussion}

\subsection{Engagement communication for robots in RLfD}

\textit{The choice of engagement cue should consider the nature of the learning task}

Our results show that robots' behavioral engagement is preferable to attentional engagement in a physical skill-oriented RLfD, which can probably be explained by the correspondence between the practice of RLfD and the cone of learning~\cite{dale1969audiovisual}.
Cone of learning, a.k.a. pyramid of learning or cone of experience, depicts the hierarchy of learning through involvement in real experiences~\cite{dale1969audiovisual}. It proposes that visual receiving (just watching the demonstration) is a passive form of learning, and learners can only remember half of the knowledge passing through this channel two weeks later. In contrast, ``doing the real thing'' is a type of active learning that leads to deeper involvement and better learning outcomes~\cite{dale1969audiovisual}.

In RLfD, the basic task for robots is to derive a policy from demonstrations and then reproduce the instructors' behavior~\cite{argall2009survey}. On the one hand, a robot's imitation behavior resemble this "behavior reproducing" process; it is thus deemed actively engaged in the learning process. 
On the other hand, although showing attentional engagement implies that the robot is involved in the visual receiving of instruction, it is still considered as a passive way to learn.
Consequently, instructors may come to the conclusion that a robot showing behavioral engagement will have deeper understanding and better mastery of the skill than that showing attentional engagement.
Moreover, by analyzing the quality gap between a robot's imitation behavior and the demonstration (behavior to be reproduced), instructors may have a more accurate assessment of the robot's learning progress.
In a word, to design effective engagement cues for robots in RLfD, we need to take the nature of the learning task into consideration. 
\\

\noindent\textit{Engagement communication should reflect robot's actual capabilities}

In our study, we do not equip the robot with any actual policy derivation algorithm since we want to avoid the perception bias caused the algorithm selection. In other words, the robot has no learning ability. Still, many subjects are convinced that robots with engagement communication (attention,  imitation, or both) would finally master the skill. They hold such a belief even if some tasks are technically very challenging for robots to learn because of the correspondence problem, e.g., swimming. These findings suggest that engagement communication can affect instructors' mental model of the robot's capability and progress. There can be a misalignment between instructors' expectations and the actual development as shown in our study. If instructors shape their teaching according to an inaccurate mental model, frustration may occur later in the RLfD process. Hence, it is critical to ensure that a robot's communication of engagement reflects its actual capabilities (policy development in the case of RLfD).
\\


%


\subsection{Limitations}
This work has several limitations. 
First, in our study, engagement communication is decoupled from the robot's actual learning process. 
However, in human or animal learning, such communication is usually associated with the learning process. 
For example, a student making good progress tends to show more behavioral engagement~\cite{skinner1993motivation}. 
We will investigate how to couple learning process with engagement communication in the future. 
Second, in this paper, we only consider two types of learning engagement cues, i.e., attention and imitation. In practice, human learners may employ more diverse cues, e.g., spatially approaching, etc. 
Third, the proposed methods, \textit{Instant attention} and \textit{Approximate imitation}, are both based on the human body poses. 
They may not be applicable to the learning tasks which do not necessarily involve the demonstrator's body movements, e.g., object manipulations. 
For those tasks, designing a good mechanism to communicate the robot engagement is still an open question. 
Fourth, in this work, we only consider skill-oriented RLfD in which the robot has to master a skill taught by instructors. 
Other types of RLfD, e.g., goal-oriented RLfD in which the robot learns how to achieve a goal from human examples, are inherently different in task settings. 
Though the proposed method may work, we still need to evaluate their effects in the future work. 
Fifth, we conduct the user study in an online simulation environment without a further off-line and real-time RLfD test. 
Though the simulation is common practice to evaluate the idea in RLfD, the participants do not have any control over the teaching process. 
How the participants might reshape future demonstration based on robot's engagement feedback needs further investigation.

\section{Conclusion and Future work}

In this work, we propose two methods (\textit{Instant attention} and \textit{Approximate imitation}) to generate robots' learning engagement in RLfD. The \textit{Instant attention} method automatically generates the point of attention and the \textit{Approximate imitation} method produces robot imitation behavior. Based on the two methods, we investigate the effects of three types of engagement communication (showing attention, showing imitation, and showing both) via a within-subject user study. Results suggest that the proposed cues enable robots to be perceived to be significantly more engaged in the learning process and behave significantly more acceptably in RLfD than with no engagement communication. 
Also, these engagement cues significantly affect the human participants' estimation of robots' learning capabilities and the participants' expectation of the learning outcomes, even though all the robots have no actual learning abilities. 
In particular, imitation cue influences instructors' perceptions significantly more than attention cue, while the hybrid cues significantly outperform a single cue. 
We also find that showing behavioral or combined engagement significantly improves instructors' assessments of demonstration qualities. 
This paper takes the first step to reveal the potential effects of communicating engagement on the humans in RLfD. 


%
%


\bibliographystyle{spmpsci}
\bibliography{myref}

\end{document}